\documentclass[showpacs,aps,amssymb,floatfix,prd,amsmath,preprintnumbers]{revtex4}
\usepackage{epstopdf}
\usepackage{capt-of}
\usepackage{graphicx}  
\usepackage{dcolumn}   
\usepackage{bm}
\begin{document}
\input epsf.tex

\title{\bf Dynamics of Bianchi $VI_h$ Universe with Bulk Viscous Fluid in Modified Gravity }

\author{B. Mishra \footnote{Department of Mathematics, Birla Institute of Technology and Science-Pilani, Hyderabad Campus, Hyderabad-500078, India, E-mail:bivudutta@yahoo.com },Sankarsan Tarai \footnote{Department of Mathematics, Birla Institute of Technology and Science-Pilani, Hyderabad Campus, Hyderabad-500078, India, E-mail:tsankarsan87@gmail.com}, S.K.J. Pacif \footnote{
Centre of Theoretical Physics, Jamia Millia Islamia, New Delhi-110025,
India, Email: shibesh.math@gmail.com}
}
\affiliation{ }

\begin{abstract}
\begin{center}
\textbf{Abstract}
\end{center}
In this paper, the dynamical behavior of the anisotropic universe has been investigated in $f(R,T)$ theory of gravity. The functional $f(R,T)$ has been rescaled in the form $f(R,T)=\mu R+\mu T$, where $R$ is the Ricci scalar and $T$ is the trace of energy momentum tensor. Three cosmological models are constructed using the power law expansion in Bianchi type $VI_h$ ($BVI_h$) universe for three different values of $h=-1,0,1$, where the matter field is considered to be of bulk viscous fluid. It is found from the model that the anisotropic $BVI_h$ model in the modified theory of gravity is in agreement with a quintessence phase for the value of $h=-1$ and $0$. We could not obtain a viable cosmological model in accordance with the present day observations for $h=1$. The bulk viscous coefficient in both the cases are found to be positive and remain constant at late time. The physical behaviours of the models along with the energy conditions are also studied.
\end{abstract}

\maketitle

\section{Introduction} 
In last five decades, the observational data provided by the cosmologists from different experiments \cite{Reiss98, Reiss99, Spergel03} gives sufficient indications that the present universe is expanding in an accelerating phase. The interesting part is that the  expansion of the universe is believed to be created from an unknown energy called dark energy (DE). This DE is about two thirds of the total energy budget of the universe. Because of the mysterious nature of the dark matter and dark energy, and the fact that their existence is inferred exclusively through their gravitational effects, it is quite natural to check whether there is a need to study these components; Specifically, whether there is any deviation from conventional General Relativity (GR) on large scales. By using the Einstein's Field Equations (EFE), there exists an accelerated expansion described by a positive constant, which is very small, in the frame work of GR, called the $\Lambda_{CDM}$ model. In the current scenario, this small positive constant is associated with dark energy in empty space, which is used to explain the recent claim of of contemporary accelerating expansion of the universe against the attractive effects of gravity.\\

Cosmologists believes that the first reason behind the possible existence of an unknown form of matter and energy driven is through the negative pressure. The second reason is the modification in the gravitational sector of the theory, which can also be considered as one of the good candidate for explaining the accelerated expansion of the universe. Some relevant alternative theories are Brans-Dicke (BD) theory, scalar-tensor theories of gravitation. Apart from these, from the physical description view, there are some other methods to express the accelerated expansion of the universe. One of these is the modified theory of gravity. Actually this theory prevents the complexities of the complicated computation of numerical solution. Another positive part of this modified theory is its consistency with recent observations for late accelerating universe and DE. Among the different models of DE, the modified gravity models are quite interesting as they incorporate some motions of quantum and general gravity theories. Some modified theories of gravity are available in the literature such as $f(R)$ gravity \cite{Nojiri06, Nojiri07, Tripathy16}, an arbitary function of Ricci scalar $R$, $f(T)$ gravity \cite{Linder10, Myrzakulov11, Chen11, Dent11, Harko14}), an arbitrary function of torsion scalar $T$ and $f(G)$ gravity \cite{Nojiri05, Li07, Kofinas14}, the Gauss-Bonnet form $G$. Among these theories $f(R)$ theory is the most appropriate theory to study the isotropic nature of the universe in FRW cosmology. The $f(T)$ theory describes the generalized version of teleparallel gravity whereas the Ricci scalar uses a more general function of the form $R=g^{ij}R_{ij}$ in $f(R)$ theory. \\

Motivated by the great success of cosmological constant as a simple and good candidate of dark energy, Harko et.al \cite{Harko11} introduced a new generalized gravity model called as $f(R,T)$ theory along the line of interest of incorporating some matter components in the action geometry. Here the Lagrangian described by an arbitrary function of the Ricci scalar $R$ and trace of the energy momentum tensor $T$. In this gravity, the cosmic acceleration may result either due to the diametrical contribution to the cosmic energy density or its dependency on matter contents. This theory can be used to examine several uses of current interest and may lead to some major differences; however of late it has been an interesting framework to investigate the accelerating models. Several authors have developed different ideas to study the nature of the universe in $f(R,T)$ gravity. \\

Belinski and Khalatnikov \cite{Belinski76} studied the viscous fluid matter in Bianchi type I space-time, They have indicated that without removing the initial big bang singularity, the viscosity can effect the qualitative aspects of the solutions around the  singularity. Many authors have studied the spatially homogeneous and anisotropic Bianchi type I space time in presence of bulk viscous by considering a constant deceleration parameter \cite{Fabris06,Saha07,Bali08,  Singh09, Sharif12a, Sharif13a, Sharif14}. Houndjo \cite{Houndjo12a} presented the cosmological reconstruction of $f(R,T)$ gravity for the rescaled functional $f(R, T)= f_{1}(R)+f_{2}(T)$ and observed the transition of matter dominated phase to the acceleration phase. Moreover Houndjo and Piattella \cite{Houndjo12b} have studied the reconstruction of the function $f(R,T)$ numerically from holographic dark energy. Sharif and Zubair \cite{Sharif12} have studied the behavior of perfect fluid and massless scalar field in $f(R,T)$ gravity for the anisotropic and homogeneous Bianchi type I space-time. Alvarenga et.al \cite{Alvarenga13} have studied the scalar density perturbations in $f(R,T)$ gravity.  Many more cosmological models are available in the literature in the framework of $f(R, T)$ gravity, either in presence of various matter distributions and space-times \cite{Moraes15, Momeni15, Jamil12, Myrzakulov12} or through energy conditions \cite{Sharif13}. \\

Mishra and Sahoo \cite{Mishra14} have obtained the exact solution of the field equations of $f(R, T)$ gravity, where the space-time is described by a Bianchi $VI_h$ space-time and the matter field is that of perfect fluid. Also Mishra et. al \cite{Mishra15} have constructed the non-static cosmological model of the universe in $f(R,T)$ gravity. Recently Mishra et al. \cite{Mishra16} have studied the dynamics of the anisotropic universe in modified gravity is studied  with a rescaled functionals. As a sequel to our previous studies on the dynamics of anistropic universe \cite{Mishra16}, in the present work, we have considered a Bianchi type $VI_h$ space time with the matter field in the form of viscous fluid. The $f(R,T)$ gravity under consideration here to study the dynamics of the universe is the generalization or modification of GR. For a specific choice of matter Lagrangian, the modified four-dimensional Einstein-Hilbert action in $f(R,T)$ gravity can be considered as 

\begin{equation} \label{eq:1}
S=\frac{1}{16\pi} \int f(R,T)\sqrt{-g}d^4x+L_m \sqrt{-g} d^4x,
\end{equation} 

where the function $f(R,T)$ can be expressed as the arbitrary function of Ricci scalar $R$ and $T=T_{ij}g^{ij}$, the trace of the stress energy tensor of matter $T_{ij}$.  $L_m$ is the matter Lagrangian density. Now, the stress energy tensor of matter is defined as 

\begin{equation} \label{eq:2}
T_{ij}=\frac{-2}{\sqrt{-g}} \frac{\delta (\sqrt{-g} L_m)}{\delta g^{ij}},
\end{equation}

Assuming that Lagrangian density of matter depends only on the metric tensor component $g_{ij}$ and not on its derivatives, therefore eqn. \eqref{eq:2} becomes

\begin{equation} \label{eq:3}
T_{ij}= g_{ij} L_m-2\frac{\partial L_m}{\partial g^{ij}}.
\end{equation}

By varying the modified four-dimensional Einstein-Hilbert action \eqref{eq:1} with respect to the metric tensor components $g^{ij}$, the algebraic function $f(R,T)$ has been chosen as a sum of two independent functions $f(R,T)=f_1(R)+f_2(T)$. $f_1(R)$ depends on the curvature $R$ whereas $f_2(T)$ is on the trace $T$ \cite{Harko11}. Hence,the generalized EFE \eqref{eq:4} yield

\begin{equation} \label{eq:4}
f_R R_{ij}-\frac{1}{2}f(R)g_{ij}+\left(g_{ij}\Box-\nabla_i \nabla_j\right)f_R = 8\pi T_{ij}+f_TT_{ij}+ \left[\bar{p} f_T+\frac{1}{2}f(T) \right] g_{ij}.
\end{equation}
$f_R=\frac{\partial f(R)}{\partial R}$ and $f_T=\frac{\partial f(T)}{\partial T}$. In order to frame a cosmological model, we assume the functional $f(R,T)$ in the form $f(R,T)= \mu R+\mu T$, subsequently the field equations \eqref{eq:4}, takes the form

\begin{equation} \label{eq:5}
R_{ij}-\frac{1}{2}Rg_{ij}=\left(1+\frac{8\pi}{\mu}\right)T_{ij}+\Lambda (T) g_{ij}.
\end{equation}
where, $\Lambda (T)=\bar{p}+\frac{1}{2}T$ can be identified with the cosmological constant which instead of being a pure constant evolves with cosmic time. 

\section{Basic Equations}
We are intending to study the dynamics of the anisotropic universe in the $f(R,T)$ gravity. We know that the standard FLRW universe is homogeneous and isotropic. Therefore, in order to address the small scale anisotropic nature of the universe, Bianchi space time is well considerable as it represents a globally hyperbolic spatially homogeneous, but not isotropic space time. So, we consider a Bianchi-type $VI_h$ space-time, where the constant exponent $h$ can be assumed values $-1,0,1$, in the form 
\begin{equation} \label{eq:6}
ds^2 = dt^2 - a_1^{2}dx^2- a^{2}_{2}e^{2x}dy^2 - a^{2}_{3}e^{2hx}dz^2,
\end{equation}
where the metric potentials $a_{i}=a_i(t); i=1,2,3$. The energy momentum tensor $T_{ij}$ for the viscous fluid can be expressed as
\begin{equation} \label{eq:7}
T_{ij}=(\rho+\bar{p})u_{i} u_{j}- \bar{p} g_{ij},
\end{equation}
where $\rho$ is the proper energy density and $\bar{p}=p-\zeta \theta$ is the viscous pressure and $\zeta$ is the bulk viscous coefficient. In the co-moving coordinate system, we have $u^{i}=(0,0,0,1).$ Also, $u^{i}=\delta^{i}_{4}$ which satisfies $g_{ij} u^{i}u^{j}=1$ and $u^{i}x_{i}=0$. With the comoving coordinate system, the field equations \eqref{eq:5} for the metric \eqref{eq:6} and energy momentum tensor \eqref{eq:7} can be obtained as,

\begin{equation} \label{eq:8}
\frac{\ddot{a_2}}{a_2}+\frac{\ddot{a_3}}{a_3}+\frac{\dot{a_2}\dot{a_3}}{a_2 a_3}- \frac{h}{a_1^{2}}= \beta\bar{p} -\frac{\rho}{2}   
\end{equation}
\begin{equation} \label{eq:9}
\frac{\ddot{a_1}}{a_1}+\frac{\ddot{a_3}}{a_3}+\frac{\dot{a_1}\dot{a_3}}{a_1 a_3}- \frac{h^2}{a_1^{2}}=\beta\bar{p} -\frac{\rho}{2}  
\end{equation}
\begin{equation} \label{eq:10}
\frac{\ddot{a_1}}{a_1}+\frac{\ddot{a_2}}{a_2}+\frac{\dot{a_1}\dot{a_2}}{a_1 a_2}- \frac{1}{a_1^{2}}=\beta\bar{p} -\frac{\rho}{2}   
\end{equation}
\begin{equation} \label{eq:11}
-\frac{\dot{a_1}\dot{a_2}}{a_1 a_2}-\frac{\dot{a_2}\dot{a_3}}{a_2 a_3}-\frac{\dot{a_3}\dot{a_1}}{a_3 a_1}+\frac{1+h+h^2}{a_1^{2}}=
\beta\rho -\frac{\bar{p}}{2}   
\end{equation}
\begin{equation} \label{eq:12}
\frac{\dot{a_2}}{a_2}+ h\frac{\dot{a_3}}{a_3}- (1+h)\frac{\dot{a_1}}{a_1}=0
\end{equation}
An over dot on the field variable denotes the differentiation with respect to time $t$ and $\beta=\left(\frac{3}{2}+\frac{8 \pi}{\mu}\right)$. In order to study the dynamical behaviour of the universe, we have redefined the above set of field equation \eqref{eq:8}- \eqref{eq:12} in the form of Hubble rates along different direction $\left(H_{1}=\frac{\dot{a_1}}{a_1}, H_{2}=\frac{\dot{a_{2}}}{a_{2}}, H_{3}=\frac{\dot{a_{3}}}{a_{3}}\right)$ as 
  
\begin{equation}\label{eq:13}
\dot{H_{2}}+\dot{H_{3}}+H^{2}_{2}+H^{2}_{3}+H_{2}H_{3}-\frac{h}{a_{1}^2}=\beta \bar{p}-\frac{\rho}{2},
\end{equation}
\begin{equation}\label{eq:14}
\dot{H_{1}}+\dot{H_{3}}+H^{2}_{1}+H^{2}_{3}+H_{1}H_{3}-\frac{h^2}{a_{1}^2}=\beta \bar{p}-\frac{\rho}{2},
\end{equation}
\begin{equation}\label{eq:15}
\dot{H_{1}}+\dot{H_{2}}+H^{2}_{1}+H^{2}_{2}+H_{1}H_{2}-\frac{1}{a_{1}^2}=\beta \bar{p}-\frac{\rho}{2},
\end{equation}
\begin{equation}\label{eq:16}
-H_{1}H_{2}-H_{2}H_{3}-H_{3}H_{1}+\frac{1+h+h^2}{a_{1}^2}=\beta \rho-\frac{\bar{p}}{2},
\end{equation}
\begin{equation}\label{eq:17}
H_{2}+hH_{3}-(1+h)H_{1}=0,
\end{equation}

The effect of both proper pressure and barotropic bulk viscous pressure can be defined as $\bar{p} = p-\zeta \theta$. From \eqref{eq:13}- \eqref{eq:17}, a general expression based on directional Hubble parameter for the effective pressure $\bar{p}$  and rest energy density $\rho$ can be established as, 
\begin{equation}\label{eq:18}
\bar{p} = p-\zeta \theta= \frac{2}{1-4 \beta^{2}}\left[\Psi(H_1,H_2,H_3,h)-2 \beta \Phi(H_1, H_2) \right] 
\end{equation}
\begin{equation} \label{eq:19}
\rho =\frac{2}{1-4 \beta^{2}}\left[ 2 \beta  \Psi(H_1,H_2,H_3,h)-\Phi(H_1, H_2)\right] 
\end{equation}

where $\Psi(H_1,H_2,H_3,h)=H_1 H_2+H_2 H_3+H_1 H_3-\frac{(1+h+h^{2})}{a_1^{2}}$ and $ \Phi(H_1, H_2)=\dot{H_1}+\dot{H_2}+H_1^{2}+H_2^{2}+H_1 H_2-\frac{1}{a_1^{2}}$. Subsequently the effective EoS parameter $\omega_{eff}=\frac{\bar p}{\rho}$ and the effective cosmological constant $\Lambda$ can be yielded from equations \eqref{eq:18} and \eqref{eq:19} as 

\begin{equation}\label{eq:20}
\omega_{eff}= 2 \beta+(1-4 \beta^{2})\frac{\Psi(H_1,H_2,H_3,h)}{2 \beta \Psi(H_1,H_2,H_3,h)-\Phi(H_1, H_2)}
\end{equation}

\begin{equation}\label{eq:21}
\Lambda=-\frac{\left[\Psi(H_1,H_2,H_3,h)+\Phi(H_1, H_2)\right]}{1+2 \beta}
\end{equation}

The bulk pressure $\bar p$, energy density $\rho$, EoS parameter $\omega_{eff}$ and effective cosmological constant (ECC) $\Lambda$ will help in investigating the dynamical behavior of the model. The understanding on the behaviour of the universe would be more appropriate if the properties of the parameters can be expressed in the form of Hubble rate. Because of the simplicity and ability to provide information about the dynamics of the universe, here we have considered the volumetric power law cosmic expansion in the form $v=t^{m}$, where $m$ is an arbitrary constant calculated from the back ground cosmology. With this assumptions, in the subsequent section, we have developed the cosmological models in $f(R,T)$ gravity for  the value of $h=-1,0,1$.

\section{Cosmological Models and Dynamical Behaviour}
Each of the value of the exponent $h$ in the filed equations leads to a cosmological model with different dynamical behaviour. 

\subsection{Case-I ($h=-1$)} 

In this case, we observed that with a suitable absorption of integrating constant with metric potential, the Hubble rate in both $y$ and $z$ direction are same i.e. $H_2=H_3$. Moreover, to study the anisotropic nature of the space time, we assumed an anisotropic relationship between the directional  Hubble parameter in the form $H_1=kH_2$, where $k$ is a positive constant. Subsequently, the  functionals $\Phi(H_1,H_2)$ and $\Psi(H_1,H_2,H_3,h)$ respectively reduced to $\Phi=(k+1)\dot{H_2}+(k^2+k+1)H_2^2-\frac{1}{a_1^2}$ and $\Psi=(2k+1)H_2^2-\frac{1}{a_1^2}$. For a power law cosmology, the directional Hubble parameters can be obtained as: $H_1=\left(\frac{km}{k+2}\right)\frac{1}{t}$, $H_2 =H_3 =\left(\frac{m}{k+2}\right)\frac{1}{t}$ and subsequently the directional scale factors provides  $a_1=t^{\frac{km}{k+2}}$ and $a_2=a_3=t^{\frac{m}{k+2}}$. So, the functionals $\Phi(H_1,H_2)$ and $\Psi(H_1,H_2,H_3,h)$ takes the form

\begin{equation}\label{eq:22}
\Phi = \left[\frac{m(1+k+k^2)}{(2+k)^2}-\frac{(1+k)}{(2+k)}\right]\frac{m}{t^2}-\frac{1}{t^{\frac{2mk}{2+k}}},
\end{equation}
\begin{equation}\label{eq:23}
\Psi = \left[\frac{(1+2k)}{(2+k)^2}\right]\frac{m^2}{t^2}-\frac{1}{t^{\frac{2mk}{2+k}}}.
\end{equation}

We know that the EoS parameter $\omega_{eff}$ and ECC $\Lambda$ as defined in \eqref{eq:20}- \eqref{eq:21} depend on the functionals $\Psi$ and $\Phi$, which are functions of the cosmic time. So, using eqns. \eqref{eq:22}-\eqref{eq:23}, in eqns.  \eqref{eq:20} and \eqref{eq:21} respectively, we obtain 

\begin{equation}\label{eq:24}
\omega_{eff}= 2 \beta+\frac{\left[\frac{(1+2k)m^{2}}{(2+k)^{2}}\right]\frac{1}{t^{2}}- t^{\frac{-2km}{2+k}}}{\left[\frac{m(1+k+k^{2})}{(2+k)^2}-\frac{(1+k)}{(2+k)}-2 \beta. \frac{(1+2k)m}{(2+k)^{2}}\right]\frac{m}{t^{2}}+(2\beta-1)t^{\frac{-2km}{2+k}}}
\end{equation}
\begin{equation}\label{eq:25}
\Lambda=\frac{1}{1+2\beta}\left[\frac{-m(1+k+k^{2})}{(2+k)^2}+\frac{(1+k)}{(2+k)}-\frac{m(1+2k)}{(2+k)^{2}}\right]\frac{m}{t^{2}}+2 t^{\frac{-2mk}{k+2}}.
\end{equation}

Since the functionals $\Psi$ and $\Phi$ are essential for analysing the EoS parameter and ECC, we are interested here to adopt a dimensional analysis on the dimensionless constant $m$ and $k$ as $m=1+\frac{2}{k}$. When $k=1$, the model reduces to an isotropic one. So, $\Phi(t)$ and $\Psi(t)$ becomes 
$\Phi(t)=\frac{(1-k^{2})}{k^{2}}\frac{1}{t^{2}}$ and $\Psi(t)=\left(\frac{1+2k-k^{2}}{k^{2}}\right)\frac{1}{t^{2}}$. Using this, we obtain from eqns. \eqref{eq:24}-\eqref{eq:25}, the corresponding EoS parameter and ECC as

\begin{equation} \label{eq:26}
\omega_{eff} = 2\beta+(1-4\beta^2)\left[\frac{k^2-2k-1}{(2\beta-1)(k^2-1)+4\beta k}\right]
\end{equation}
\begin{equation}\label{eq:27}
\Lambda(t)= \frac{2}{1+2\beta}\left[1-\frac{k+1}{k^2}\right]\frac{1}{t^2}.
\end{equation}

So, from eqn. \eqref{eq:26}, we can infer that for a given value of scaling constant $\mu$, $\left(\beta=\frac{3}{2}+\frac{8 \pi}{\mu}\right)$, the EoS parameter is constant as the anisotropic parameter $k$ is also a constant. It is also observed from eqn.\eqref{eq:27} that the ECC decreases quadratically with the increase in cosmic time; of course with a given scaling constant. To frame a realistic cosmological model, we need to address the scaling constant and anisotropic parameter in such a way that the EoS parameter would be negative and would be less than $-1/3$ at late times. Moreover, in order to achieve a viable cosmological model, the ECC should be large at initial time and should vanish at late times. The same has been represented in Fig-1 and Fig-2.

\begin{figure}[h!]
\minipage{0.40\textwidth}
\centering
\includegraphics[width=7.2cm,height=7.5cm]{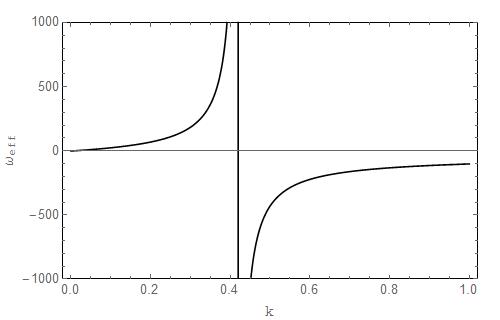}
\caption{ $\omega_{eff}$ versus $k$ for $h=-1$.}
\label{fig1}
\endminipage\hfill
\minipage{0.40\textwidth}
\includegraphics[width=7.2cm,height=7.5cm]{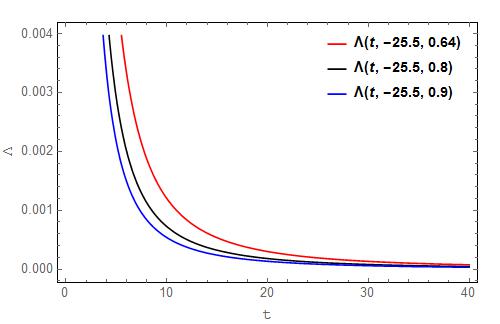}
\caption{ $\Lambda$ versus $t$ for $h=-1$.}
\label{fig2}
\endminipage

\end{figure}
 
From Fig-1, as indicated earlier, we have observed that $\omega_{eff}$ is a constant value for a given value of $\beta$ and assumed anisotropy parameter $k$. According to the observational data the $\omega_{eff}$ should stay in negative axis and less than $-\frac{1}{3}$. To stay in negative axis we chose  a negative value of the model parameter $(\beta =-25.5).$ We can observed from the figure that the EoS parameter $\omega_{eff}$ increases nearly from a negative value for lower value of $k$ to zero at higher value of $k$. As indicated earlier, in Fig-2, the ECC varies from large positive values in early epoch  to almost vanished at late time.

\subsection{Case-II ($h=0$)}

Substituting the value of the exponent $h=0$ in eqn. \eqref{eq:17}, we observed that the Hubble rate is same both in $x$ and $y$ directions. With an assumed anisotropy relation on the $y$ and $z$ direction in the form $H_{3}=r H_{2} $ leads the directional Hubble rates in power law expansion of volume scale factor as $ H_{1}=H_{2}=\left(\frac{m}{r+2}\right)\frac{1}{t} $ and $ H_{3}=\left(\frac{mr}{r+2}\right)\frac{1}{t}$. Thus the corresponding metric potentials are $a_1=a_2=t^{\frac{m}{r+2}}$ and $ a_3=t^{\frac{mr}{r+2}}$. The functionals $ \Phi(t) $ and $ \Psi(t) $ for this model are 

\begin{equation} \label{eq:28}
\Phi(t) = \left[\frac{3m^2-2m(r+2)}{(r+2)^2}\right]\frac{1}{t^2}-\frac{1}{t^{\frac{2m}{r+2}}}
\end{equation}

\begin{equation} \label{eq:29}
\Psi (t) = \left[\frac{(2r+1)m^2}{(r+2)^2}\right]\frac{1}{t^2}-\frac{1}{t^{\frac{2m}{r+2}}}.
\end{equation}

As in the previous case, here also we have employed the dimensional consistency term $m=r+2$, in eqns. \eqref{eq:28} and \eqref{eq:29}. As a result, $\omega_{eff}=\frac{1}{2\beta}$, which is a constant and the ECC, $\Lambda=-\frac{2r}{(2\beta+1)t^2}$, is time varying. Again, to obtain a viable cosmological model, the scaling constant has been constrained to be negative, which ultimately assumed $\beta$ to be negative. With this constraint, $\omega_{eff}$ would be in the negative domain and do not affect by the choice of anisotropy in the model, as there is no anisotropy term in the expression. The same has been depicted in Fig-3. The ECC remains in the positive domain and decreases with increase in time (Fig-4). It is important to note here that $\omega_{eff}$ lies in the quintessence region when the scaling parameter is $\leq-\frac{1}{2}$ and when it is more that $\frac{1}{2}$ enters into the phantom region.    

\begin{figure}[h!]
\minipage{0.40\textwidth}
\centering
\includegraphics[width=7.2cm,height=7.5cm]{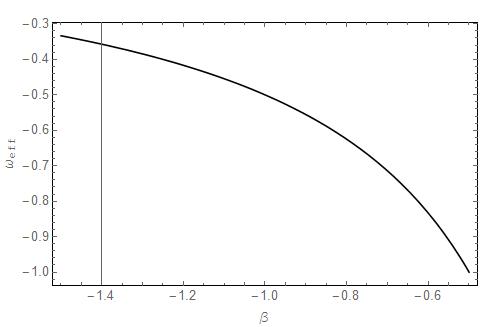}
\caption{$\omega_{eff}$ versus $\beta$ for $h=0$.}
\label{fig3}
\endminipage\hfill
\minipage{0.40\textwidth}
\includegraphics[width=7.2cm,height=7.5cm]{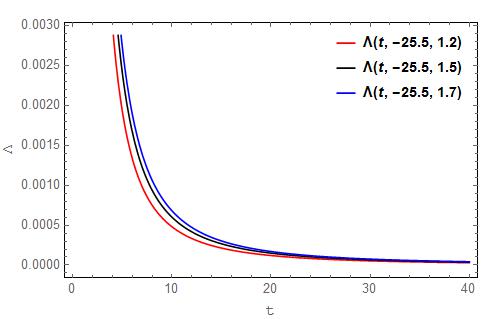}
\caption{$\Lambda$ versus $t$ for $h=0$.}
\label{fig4}
\endminipage
\end{figure}

\subsection{Case-III ($h=1$)}

In this case, substituting $h=1$, again in \eqref{eq:17}, we found that the change of  Hubble rate in $x-$ direction is half of the sum of the Hubble rate in $y-$ and $z-$ directions. This leads to another important fact that the mean Hubble rate and the Hubble rate in $x-$ direction are same. As in the preceding section, here also, we have used the power law cosmology in the form $v=t^m$ and obtained the functional $\Phi$ and $\Psi$ as 

\begin{equation} \label{eq:30}
\Phi(t) = 2\dot{H}+3H^2+\frac{\tau^2}{t^{2m}}-\frac{1}{t^{\frac{2m}{3}}},
\end{equation}

\begin{equation} \label{eq:31}
\Psi(t)= 3H^2-\frac{\tau^2}{t^{2m}}-\frac{3}{t^{\frac{2m}{3}}}.
\end{equation}

Where $\tau$ is an integrating constant. In order to make the functionals dimensional consistent, the value of the exponent $m$ should be 3. With this value of the exponent $m$, the deceleration parameter would not be negative, which in turn does not provide an accelerating model. Moreover, $w_{eff}$ found to be unity, which is not in agreement with  the dark energy driven cosmic acceleration; though the ECC vanishes. Therefore Bianchi type $VI_h (h=1)$ space-time is not compatible in the study of present day accelerated expansion of the universe.

\section{Physical Parameters and Energy Conditions}

In this section, we have analysed the behaviour of the physical parameters of the cosmological models obtained earlier. The power law model studied  based on the fact that the growth of the scale factor ($a(t)\propto t^{m})$ depends on the exponent $m$. When $m$ lies in the positive domain the observed universe is expanding whereas it contracts for a negative $m$. We know that the role of  Hubble parameter and the deceleration parameter inscribed in the study of power law cosmology. We obtained both the parameters in the form $H=\frac{1}{3}\left(\frac{\dot{a_1}}{a_1}+\frac{\dot{a_2}}{a_2}+\frac{\dot{a_3}}{a_3}\right)=\frac{m}{3t}$ and $q=-\frac{\ddot{a}a}{\dot{a}^{2}} =-1+\frac{3}{m}$. From the deceleration parameter it is quite clear that in order to have a viable cosmological model, the value of $m$ must be less than 3. So, the Hubble parameter decreases with increase in time and may vanish at infinite future. The scalar expansion of the model is $\theta=\Sigma H_i=\frac{m}{t}$, which also indicates that it decreases with time and may vanish at late time. The shear scalar of the model, $\sigma^{2}=\frac{m^{2}}{3 t^{2}}$, and the average anisotropy parameter $\mathcal{A}$ is defined to be $\mathcal{A}=\frac{1}{3}\Sigma \left(\frac{\Delta H_i}{H}\right)^{2}$.
The viscous coefficient $\xi$ for  $h=-1$ and for $h=0$  can be respectively calculated as 

\begin{equation}
\xi=\frac{2}{4\beta^2-1}\left[\frac{(\gamma-2\beta)(1-k^2)+(1-2\beta\gamma)(1+2k-k^2)}{k(k+2)}\right]\frac{1}{t}
\end{equation}  

\begin{equation}
\xi=\frac{2}{4\beta^2-1}\left[\frac{(1-2\beta r)2r}{r+2}\right]\frac{1}{t}
\end{equation} 

\begin{figure}[h!]
\minipage{0.40\textwidth}
\centering
\includegraphics[width=7.2cm,height=7.5cm]{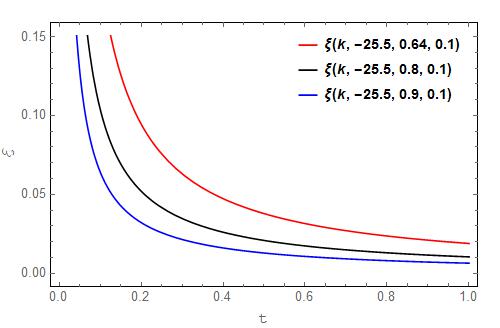}
\caption{$\xi$ versus cosmic time $t$ for $h=-1$.}
\label{fig3}
\endminipage\hfill
\minipage{0.40\textwidth}
\includegraphics[width=7.2cm,height=7.5cm]{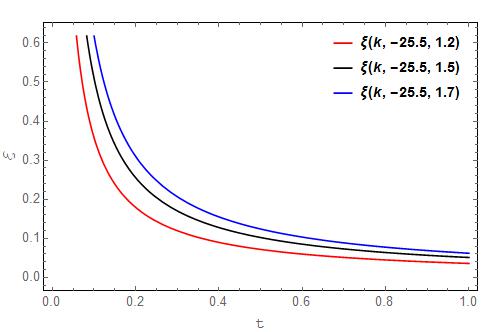}
\caption{$\xi$ versus cosmic time $t$ for $h=0$.}
\label{fig4}
\endminipage
\end{figure}

Fig-5 and Fig-6 respectively gives the graphical representation of the viscous coefficient. In both cases it is observed that the coefficient remains positive throughout. Even if, for different representative value of the anisotropy parameter $k=0.64,0.8,0.9$ in Fig-5 and $r=1.2,1.5,1.7$ in fig 6, the coefficient behave same. It is also observed that in both the cases the bulk viscous coefficient remain constant throughout. The state finder diagnostic pair that gives an impression on the geometrical nature of the model is found to be $r=\left(1-\frac{3}{m}\right)\left(1-\frac{6}{m}\right)$ and $s=\frac{2}{m}$. For a large value of the anisotropy relation $m$, the state finder pair having value $(1,0)$. \\

The idea of energy condition came from the famous Raychoudhuri equations \cite{Raychoudhuri55,Raychoudhuri79} which play a key role to discuss the congruence of null and time like geodesics with the requirement that not only the gravity is attractive but also the energy density is positive. The energy conditions are few additional conditions for the matter content of the gravitational theory. Energy conditions are the co-ordinate invariant which incorporate the common characteristics shared by almost every matter field. In GR, the role of these energy conditions is to prove the theorems about the existence of space time singularity and black holes \cite{Wald84}. For perfect fluid matter distribution, these inequalities provide the energy constraints defined by: Null Energy Condition (NEC):$ \rho+p\geqslant0$, Weak Energy Condition (WEC): $\rho\geqslant0$ ,  $\rho+p\geqslant0 $, Dominant Energy Condition (DEC):$\rho-p\geqslant0$,  $\rho+p\geqslant0$, Strong Energy Condition (SEC): $\rho+p\geqslant0$, $\rho+3p\geqslant0$. These condition shows that if NEC violates, the other energy conditions also violates. \\

For the case $(h=-1)$,with the help eqns. \eqref{eq:18}-\eqref{eq:19}, we obtained different energy conditions as: 
\begin{equation} \label{eq:34}
NEC:\rho+p=\frac{2}{4\beta^{2}-1}\left[\left(\frac{(1-2\beta)(1-k^{2})-4\beta k}{k^{2}}\right) (1+\gamma)\right]\frac{1}{t^{2}}\geqslant0
\end{equation}

\begin{equation} \label{eq:35}
WEC: \rho=\frac{2}{4\beta^{2}-1}\left[\frac{(1-2\beta)(1-k^{2})-4\beta k}{k^{2}}\right]\frac{1}{t^{2}}\geqslant0
\end{equation}

\begin{equation} \label{eq:36}
SEC: \rho+3p=\frac{2}{4\beta^{2}-1}\left[\left(\frac{(1-2\beta)(1-k^{2})-4\beta k}{k^{2}}\right) (1+3\gamma)\right]\frac{1}{t^{2}}\geqslant0
\end{equation}

\begin{equation} \label{eq:37}
DEC:\rho-p=\frac{2}{4\beta^{2}-1}\left[\left(\frac{(1-2\beta)(1-k^{2})-4\beta k}{k^{2}}\right) (1-\gamma)\right]\frac{1}{t^{2}}\geqslant0
\end{equation}
\begin{figure}[h!]
\minipage{0.40\textwidth}
\centering

\includegraphics[width=7.2cm,height=7.5cm]{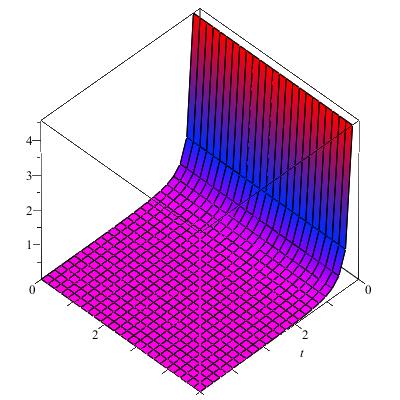}
\caption{Null energy condition (NEC) for $\beta=-25.5$ ,$k=0.64$ and $\gamma=0.1.$  }
\label{fig4}
\includegraphics[width=7.2cm,height=7.5cm]{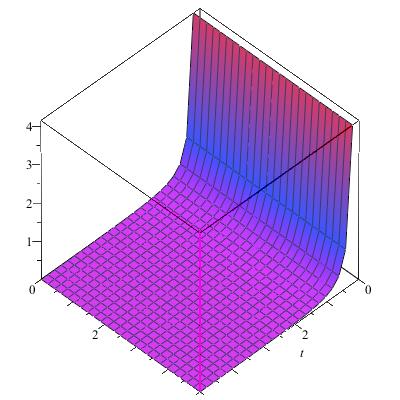}
\caption{Weak energy condition (WEC) for $\beta=-25.5$, $k=0.64$ and $\gamma=0.1.$}
\label{fig5}
\endminipage\hfill
\minipage{0.40\textwidth}
\includegraphics[width=7.9cm,height=7.5cm]{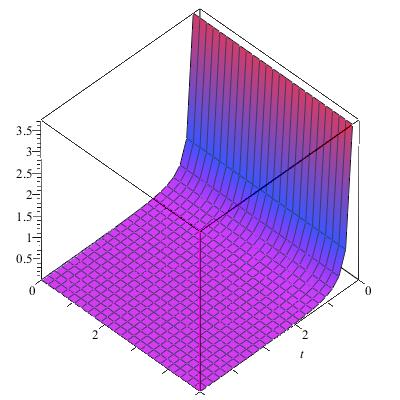}
\caption{Dominant energy condition(DEC) for $\beta=-25.5$, $k=0.64$ and $\gamma=0.1$}
\label{fig6}
\includegraphics[width=7.9cm,height=7.5cm]{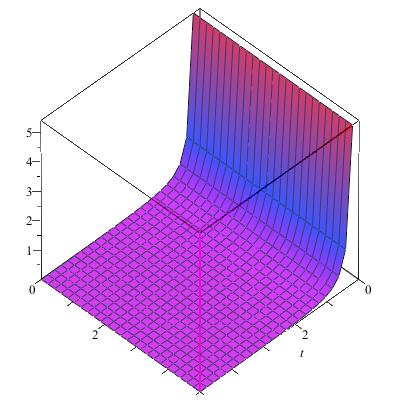}
\caption{Strong energy condition(SEC) for $\beta=-25.5$, $k=0.64$ and $\gamma=0.1$}
\label{fig7}
\endminipage
\end{figure}
We have analysed the energy conditions through graphical representation with the representative value of the parameter $\beta=-25.5$, $k=0.64$ and $\gamma=0.1$. Fig-7  and Fig-8, the behaviour of NEC and WEC to the  present model has been shown. In these figures, we surveyed that NEC as well as WEC are dynamically varying from a large positive values at initial stage to vanishingly null values at late time of cosmic evolution with positive axis. Fig-9 and Fig-10, SEC and DEC show the similar behaviour with respect to time. These graphs show that for large value of time the model is decreasing at initial stage to vanishingly null values at late time. From the above relation we found that all conditions are satisfying this model and remain attractive for gravity through energy momentum tensor.\\

Similarly for $h=0$, we have obtained the energy conditions as:

\begin{equation} \label{eq:38}
NEC:\rho+p=\frac{-8 \beta n (1+\gamma)}{(4 \beta^{2}-1)t^{2}}\geqslant0
\end{equation}

\begin{equation} \label{eq:39}
WEC:=\frac{-8 \beta n \gamma}{(4 \beta^{2}-1)t^{2}}\geqslant0
\end{equation}

\begin{equation} \label{eq:40}
SEC:\rho+3p=\frac{-8 \beta n (1+3\gamma)}{(4 \beta^{2}-1)t^{2}}\geqslant0
\end{equation}

\begin{equation} \label{eq:41}
DEC:\rho+p=\frac{-8 \beta n (1-\gamma)}{(4 \beta^{2}-1)t^{2}}\geqslant0
\end{equation}

\begin{figure}[h!]
\minipage{0.40\textwidth}
\centering
\includegraphics[width=7.2cm,height=7.5cm]{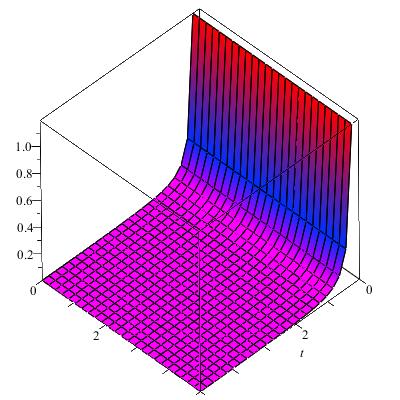}
\caption{NEC for $\beta=-25.5$, $k=1.2$ and $\gamma=0.1$ }
\label{fig4}
\includegraphics[width=7.2cm,height=7.5cm]{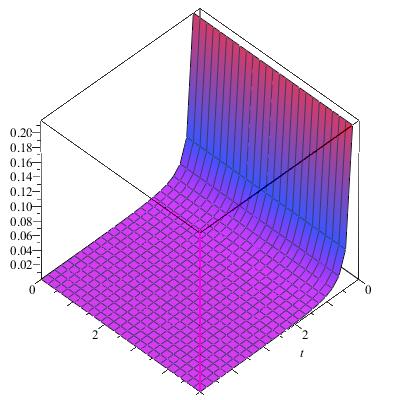}
\caption{WEC for  $\beta=-25.5$, $k=1.2$ and $\gamma=0.1$}
\endminipage\hfill
\minipage{0.40\textwidth}
\includegraphics[width=7.2cm,height=7.5cm]{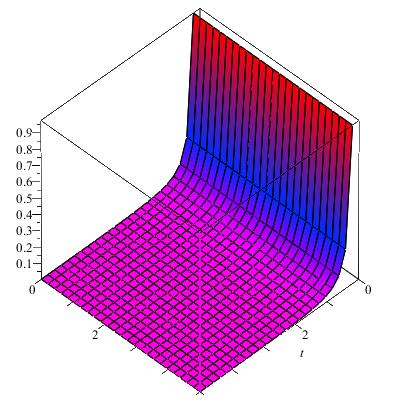}
\caption{DEC for  $\beta=-25.5$, $k=1.2$ and $\gamma=0.1$}
\label{fig5}
\includegraphics[width=7.2cm,height=7.5cm]{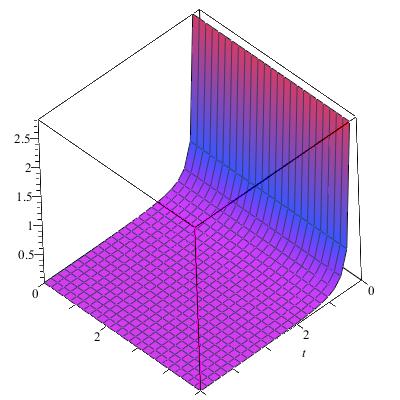}
\caption{SEC for  $\beta=-25.5$, $k=1.2$ and $\gamma=0.1$}
\label{fig6}
\endminipage
\end{figure}
In Fig-11 to Fig-14, we have plotted different energy conditions for this model with the respective value of $\beta$, $k$ and $\gamma.$ From the mathematical perspective, recent observational data suggest that the SEC is violated by matter configuration. As per the recent observations, the researchers are preferred to work with NEC, as it is easy to analyse the behaviour of universe. In this model the NEC and SEC behaves the same behaviour as shown in the previous model.
\section{Conclusion}
In view of the growing interest in modified theory of gravity, we have constructed the cosmological models of the universe in $f(R,T)$ gravity keeping the dimensional consistency at the background. The linear functional $f(R)=\mu R$ and $f(T)=\mu T$ considered here generates the idea of a time varying ECC. For $h=-1$ anf $h=0$, we could find a viable cosmological models; however for $h=1$, a viable cosmological model could not be obtained. In the first two cases, where the viable cosmological models were obtained, the ECC start evolving from large positive value initially and subsequently become small at late times. This result is in accordance with the present observations on dark energy driven cosmic acceleration. For a large $m$, the state finder diagnostic pair having the value $(1,0)$, which is in agreement with the behavior of $\Lambda_{CDM}$ model.

\section{Acknowledgement}
BM acknowledges SERB-DST, New Delhi, India for financial support to carry out the Research project[No. SR/S4/MS:815/13].


\begin{thebibliography}{99}

\bibitem{Reiss98} A.G. Riess et al., \textit{Astronomical Journal}, \textbf{116(3)}, 1009, (1998).

\bibitem{Reiss99} A.G. Riess et al.,  \textit{Astronomical Journal}, \textbf{117(3)},  707,(1999).

\bibitem{Spergel03} D.N. Spergel et al., \textit{Astrophys. J. Suppl}, \textbf{148}, 175,  (2003).

\bibitem{Nojiri06} S. Nojiri, S.D. Odintsov, \textit{ Physical Review D}, \textbf{74(8)}, 086005, (2006).

\bibitem{Nojiri07} S. Nojiri, S.D. Odintsov, \textit{Physics Letters B}, \textbf{657(4)}, 238, (2007).

\bibitem{Tripathy16} S.K. Tripathy, B.Mishra, \textit{The European Physical Journal Plus}, textbf{131}, 273, (2016).

\bibitem{Linder10} E.V. Linder, \textit{Physical Review D} , \textbf{81}, 127301, (2010).

\bibitem{Myrzakulov11} R. Myrzakulov, \textit{The European Physical Journal C},\textbf{ 71(9)}, 1752, (2011).

\bibitem{Chen11} S.H. Chen, J.B. Dent, S. Dutta, E.N. Saridakis, \textit{Physical Review D}, \textbf{83(2)}, 023508, (2011).

\bibitem{Dent11} J.B. Dent, S. Dutta, E.N Saridakis, \textit{Journal of Cosmology and Astroparticle Physics} , \textbf{2011},009, (2011).

\bibitem{Harko14} T.Harko, F.S. Lobo, G. Otalora, E.N. Saridakis,  \textit{Physical Review D}, \textbf{89(12)}, 124036, (2014).

\bibitem{Nojiri05} S. Nojiri, S.D. Odintsov,\textit{Physics Letters B} ,\textbf{631(1)}, 1, (2005).

\bibitem{Li07} B. Li, J.D. Barrow, D.F. Mota, D. F., \textit{Physics Letters B} , \textbf{76(4)}, 044027, (2007).

\bibitem{Kofinas14} G. Kofinas, E.N. Saridakis, \textit{Physical Review D}, \textbf{90}, 084044,(2014).

\bibitem{Harko11} T. Harko, F.S.N. Lobo, S. Nojiri, S.D. Odintsov, \textit{ Phys. Rev. D}, \textbf{84}, 024020, (2011).

\bibitem{Belinski76} V. A. Belinski, I. M. Khalatnikov, \textit{Sov. Phys. JETP}, \textbf{42}, 205, (1976).

\bibitem{Fabris06} J. C. Fabris, S. V. B. Goncalves, R. S. Rebeiro, \textit{Gen. Relativ. Gravit.}, \textbf{38}, 495,(2006).

\bibitem{Saha07} B. Saha, \textit{Astrophys. Space Sci.}, \textbf{312},3, (2007).

\bibitem{Bali08} R. Bali, J. P. Singh, \textit{Int. J. Theor. Phys.}, \textbf{47}, 3288, (2008).

\bibitem{Singh09} C. P. Singh, S. Kumar, \textit{Int. J. Theor. Phys.}, \textbf{48}, 925, (2009).
\bibitem{Sharif12a} M. Sharif, M. Zubair, \textit{Astrophys Space Sci},\textbf{339}, 45, (2012).

\bibitem{Sharif13a} M. Sharif, M. Zubair, \textit{J. Phys. Soc. Jpn.}, \textbf{82}, 014002, (2013).

\bibitem{Sharif14}M. Sharif, M. Zubair, \textit{Astrophysics and Space Science},\textbf{349}, 457, (2014).

\bibitem{Houndjo12a} M.J.S. Houndjo, \textit{Int. J. Mod. Phys. D}, \textbf{21}, 1250003, (2012).

\bibitem{Houndjo12b} M.J.S. Houndjo, O.F. Piattella, \textit{Int. J. Mod. Phys. D }, \textbf{21}, 1250024 (2012).

\bibitem{Sharif12} M. Sharif, M.J. Zubair, \textit{ Phys.Soc. Jpn.},\textbf{81}, 114005, (2012).

\bibitem{Alvarenga13} Alvarenga et.al, \textit{Physical Review D}, \textbf{87}, 103526,(2013).

\bibitem{Jamil12} M. Jamil, D. Momeni, R. Myrzakulov, \textit{Chin. Phys. Lett.}, \textbf{29} ,109801 (2012).

\bibitem{Myrzakulov12} R. Myrzakulov, \textit{Eur. Phys. J. C} \textbf{72}, 2203 (2012).

\bibitem{Moraes15} P.H.R.S. Moraes, \textit{Eur. Phys. J. C}, \textbf{75}, 168 (2015).

\bibitem{Momeni15} D. Momeni, R. Myrzakulov, E. Gudekli, \textit{Int. J. Geom. Methods Mod. Phys.}, \textbf{12}, 1550101 (2015).

\bibitem{Sharif13} M. Sharif, et al., \textit{Eur. Phys. J. Plus}, \textbf{128}, 123 (2013).

\bibitem{Mishra14} B. Mishra, P.K. Sahoo, \textit{Astrophys. Space Sci.}, \textbf{352(1)}, 331, (2014).

\bibitem{Mishra15} B. Mishra, P.K. Sahoo, S. Tarai, \textit{ Astrophys Space Sci.}, \textbf{359}, 15, (2015).

\bibitem{Mishra16} B. Mishra, S. Tarai, S.K. Tripathy\textit{Adv.High Energy Phys.}, textbf{  8543560},1,(2016).

\bibitem{Raychoudhuri55} A. Raychoudhuri, \textit{Phys. Rev}, textbf{98}, 1123, (1955).

\bibitem{Raychoudhuri79} A. Raychoudhuri, \textit{Theoretical Cosmology(Oxford University Press)}, (1979).

\bibitem{Wald84} R. M. Wald, \textit{General Relativity (Chicago, IL: University of Chicago Press)} (1984).




\end{thebibliography}
\end{document}